\documentclass[pra,twocolumn,notitlepage,longbibliography]{revtex4-2}
\usepackage{amsmath}
\usepackage{amssymb}
\usepackage{bbold}
\usepackage[unicode=true,colorlinks=true,citecolor=blue,urlcolor=blue]{hyperref}
\newcommand{\ket}[1]{|#1\rangle}
\newcommand{\bra}[1]{\langle #1|}
\usepackage{bm}
\usepackage{epsfig}

\usepackage[normalem]{ulem}

\renewcommand {\Im}{\mathop\mathrm{Im}\nolimits}

\renewcommand {\i}{{\rm i}}
\renewcommand {\phi}{{\varphi}}
\newcommand {\rmi}{{\rm i}}

\begin{document}
\title{
Optomechanical amplification driven by interference of phonon-exciton and phonon-photon couplings
}

\author{Egor S. Vyatkin}
\affiliation{Ioffe Institute, St. Petersburg 194021, Russia}

\author{Alexander N. Poddubny}
\email{poddubny@coherent.ioffe.ru}

\affiliation{Ioffe Institute, St. Petersburg 194021, Russia}

\begin{abstract}
We study theoretically optomechanical damping and amplification spectra for vibrations interacting with excitonic polaritons in a zero-dimensional microcavity. We demonstrate, that the spectra strongly depend on the ratio of the exciton-phonon and the photon-phonon coupling constants. The interference between these couplings enables a situation when optomechanical gain exists either only for a lower polaritonic resonance or only for an upper polaritonic resonance. Our results provide  insight in the optomechanical interactions in various multi-mode systems, where several resonant oscillators, such as photons, plasmons, or excitons are coupled to the same vibration mode.
\end{abstract}
\date{\today}

\maketitle
\section{Introduction}\label{sec:intro}

Cavity optomechanics of nanoscale semiconductor structures is now rapidly developing~\cite{Delsing_2019}.  Such  paradigmatic effects as phonon lasing~\cite{Grudinin2010,Maryam2013}, optomechanical heating and cooling~\cite{Xie2021} and optomechanical nonreciprocity \cite{Sohn2018} have already been  demonstrated.  However, the quest for compact on-chip structures with strong optomechanical interactions and long coherence times is still ongoing. One of  important recent milestones in this field is a demonstration of Bragg  microcavity, trapping both light and sound with record-high sound frequency of $\approx 0.1$~THz~\cite{Fainstein2017}. Even more opportunities are offered by  optomechanical semiconductor microcavities with embedded quantum wells or quantum dots~\cite{kavbamalas}. In such structures resonant interaction of light with localized excitons leads to formation of excitonic polaritons, that are hybrid half light-half matter quasiparticles. Polaritonic platform has a number of potential advantages. Contrary to the traditional setup where vibrations are coupled only to a single localized phonon mode, here  vibrations interact  both with light and excitons~\cite{Baker2014}. Namely, exciton resonance frequency  shifts due to the deformation potential, that leads to the photoelastic mechanism of optomechanical coupling and  can significantly increase the  coupling strength~\cite{Jusserand2015}.  It has been predicted that the polaritonic microcavities can manifest dissipative \cite{Vishnevsky2011,Kyriienko2014} and nonlinear interactive optomechanical coupling, with the latter being very sensitive to polariton-polariton interactions~\cite{Bobrovska2017}. The phonon lasing mediated by Bose condensate of the excitonic polaritons has been demonstrated very recently \cite{Chafatinos2020}. The advantage of such Bragg microcavity platform is the compatibility with the planar technology, making it relatively easy to integrate several resonators. For example, interaction of several polaritons, trapped in different points of the Bragg microcavity  plane, with the  same surface acoustic mode has been achieved \cite{Kuznetsov2020}.

Despite this rapid experimental progress, details of the optomechanical interactions in a polaritonic system, are still not yet fully understood. In this work we study theoretically the optomechanical gain and loss in a very simple model, where a single vibration mode interacts with both exciton and photon modes at the same time with the coupling strengths $g_b$ and $g_c$, respectively, see the schematic illustration in Fig.~\ref{fig:1}.
Naively, one could independently consider  the coupling of vibrations to the  upper and the lower polariton mode. Here we show, however, that the interference between optomechanical couplings for upper and lower polaritonic branches plays an important role and thus the problem can not be reduced to the two independent ones. This result holds even in the strong coupling regime, where the Rabi splitting $g$ between the two polaritonic resonances is much larger than their linewidths. The key parameter, controlling whether the interaction of vibrations with the two polaritonic states can be considered as indepedent one, is the ratio of the phonon frequency $\Omega$ to the Rabi splitting~\cite{Wu_2013}. We consider both the linearized approximation, valid for low optical pumping strength and the strongly nonlinear phonon lasing regime. 

To the best of our knowledge, the phonon lasing for a single vibration mode interacting with two resonances at the same time has been studied only for a specific case $g_c/g_b=-1$, when the two optomechanical coupling constants have equal magnitude and opposite sign~\cite{Wu_2013}. Here, we demonstrate that the optomechanical gain and the attractors of nonlinear oscillation are very sensitive to the ratio of the couplings $g_c/g_b$. By tuning the value of $g_c/g_b$ it is possible to realize a situation when optomechanical gain exists either only for lower polaritonic resonance or only for upper polaritonic resonance.

Our findings of tunable optomechanical gain are potentially applicable not only for semiconductor polaritonic platform but also for other systems where vibration mode interacts with several resonances.  For example, a new optomechanical platform of a molecule coupled to the localized plasmon mode has been put forward in Refs.~\cite{Benz2016,Lombardi2018} and it it could be possible also to realize a situation when there  two plasmonic resonances interacting with the molecule at the same time.

The rest of the paper is organized as follows. We start with a more detailed discussion of potential platforms to observe the optomechanical interference in Sec.~\ref{sec:realizations}. The  calculation approach is discussed in Sec.~\ref{sec:model}. Section \ref{sec:results} presents the optomechanical gain spectra for weak oscilation amplitude when  the linearized approximation is applicable. The nonlinear regime for larger amplitudes is discussed in Sec.~\ref{sec:nonlinear}. The reults are summarized in Sec.~\ref{sec:summary}.















\section{Potential realizations}\label{sec:realizations}
\begin{figure}[t!]
\includegraphics[width=0.4\textwidth]{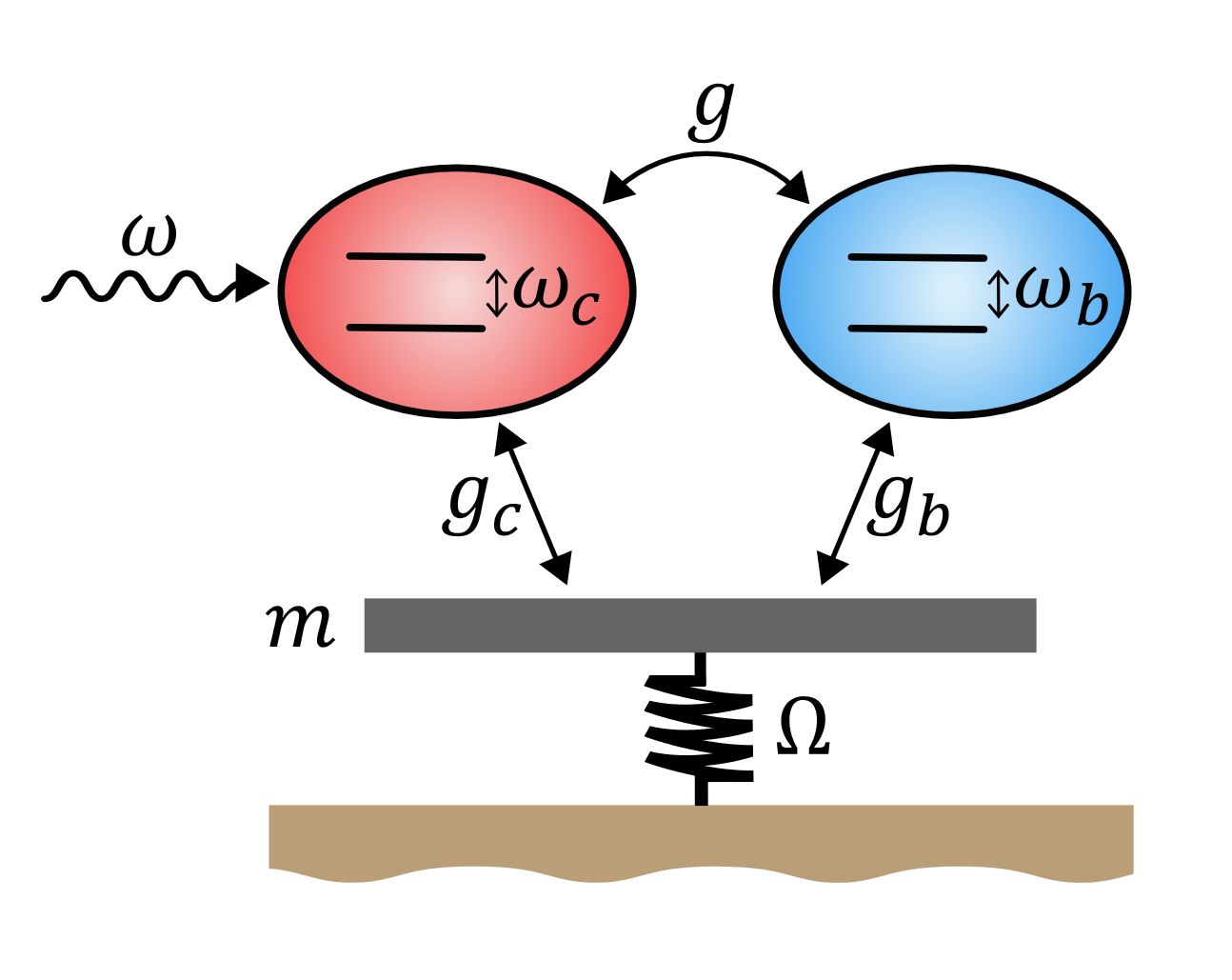}
\caption{Scheme of the considered system of coupled cavity and  exciton resonances $
\omega_c$ and $
\omega_b$ interacting with the common vibration mode with the mechanical frequency $\Omega$.}\label{fig:1}
\end{figure}

\begin{figure}[t]
\centering\includegraphics[width=0.4\textwidth]{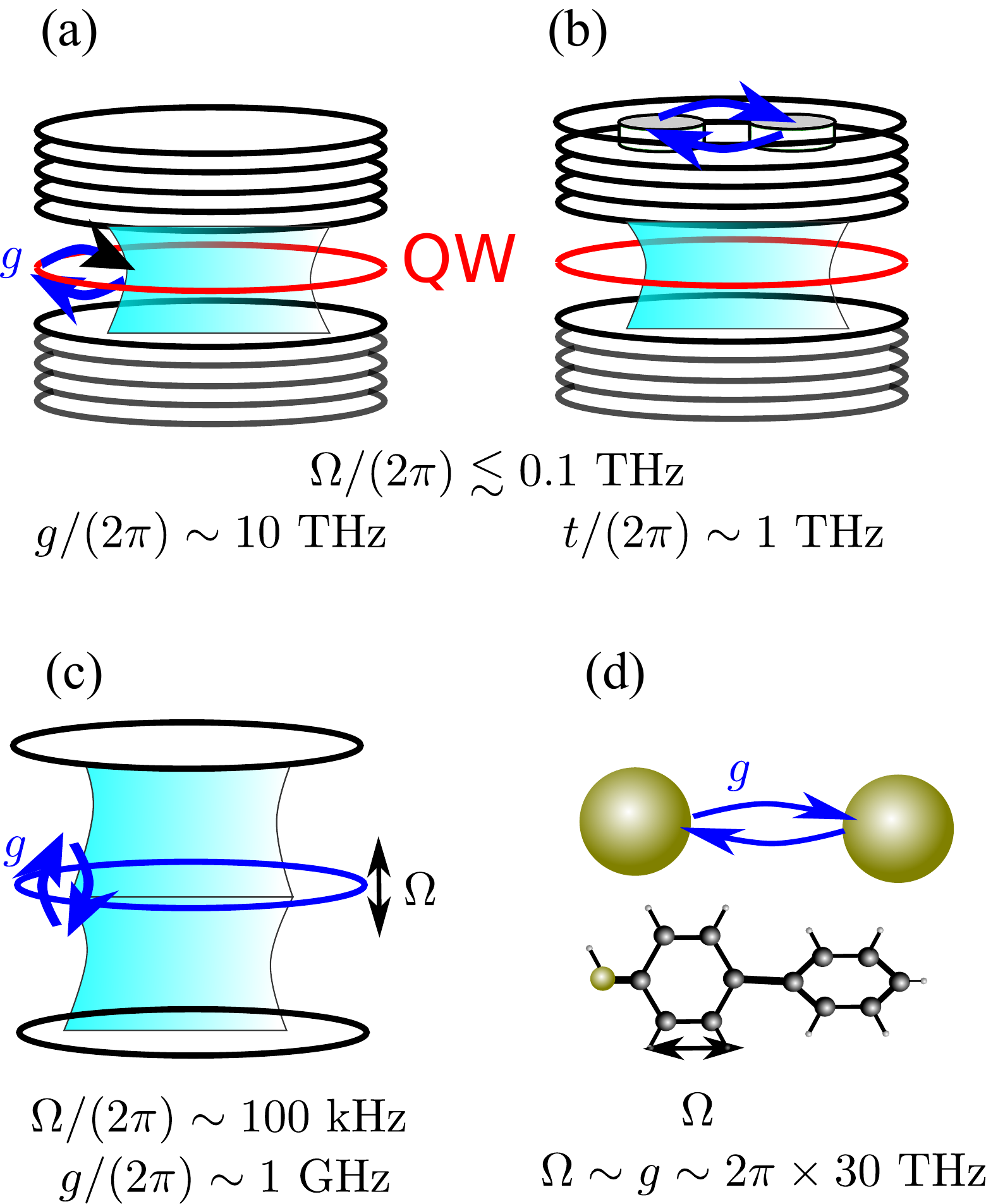}
\caption{Schematics of several optomechanical structures with two coupled resonances: (a) Bragg polaritonic micropillar cavity with an embedded quantum well (QW) from \cite{Chafatinos2020} (b) Same as (a) but with polariton traps introduced by mirror patterning (c)  Two Fabry-P\'erot cavities separated by a micromechanical membrane \cite{Sankey2010} (d) Molecular picocavity with metallic nanoparticles, corresponding to \cite{Benz2016,Lombardi2018}.
Typical phonon frequency $\Omega$ and the strength  of the coupling $g$ ($t$) between the two resonances are indicated.
}\label{fig:structures}
\end{figure}
The generic scheme for the considered optomechanical structure, where  vibrations interact with two coupled resonances, is shown in Fig.~\ref{fig:1}. Before presenting the theoretical details of the model and calculation scheme, it is intructive to discuss potential experimental realizations of this setup, illustrated in Fig.~\ref{fig:structures}. Figure~\ref{fig:structures}(a) shows the Bragg micropillar cavity~\cite{Fainstein2017,Chafatinos2020} with an embedded quantum well (QW) that traps phonons and light at the same time~\cite{Jusserand2013}. While the frequency of confined phonons  in such a structure can be quite large, it is still smaller than the typical vacuum Rabi splitting ~\cite{Khitrova2006} between photons and excitons by two orders of magnitude, $\Omega/g\sim 10^{-2}$. As will be demonstrated below, see also Ref.~\cite{Wu_2013}, for such large Rabi splitting the coupling of vibrations  to upper and lower polaritonic resonances is quasi-independent and interferences of two optomechanical couplings that we focus on in this work, are suppressed. However, novel opportunities are opened for patterned microcavities  \cite{Chafatinos2020,Kuznetsov2018}. Specifically, as shown in Fig.~\ref{fig:structures}(b) one can etch the upper Bragg mirror and induce  traps for lower branch polaritons. The two-trap system is described by the same scheme Fig.~\ref{fig:1}. The only difference is that  the exciton and photon resonances in Fig.~\ref{fig:1} are replaced by two resonances of trapped polaritons and the coupling strength $g$ is replaced by the effective tunneling constant between the traps $t$ that can be on the order of $1$~THz, i.e. not much larger than the cavity frequency $\Omega$. The potential problem with this setup could be the difficulty to tune the couplings $g_c$ and $g_b$ independently since they will be determined by the overlap of the polaritons with the same confined phonon mode. It seems especially challenging  to realize the situation when $g_c/g_b<0$.

Figure~\ref{fig:structures}(c) presents the original setup from \cite{Sankey2010},  when the two Fabry-P\'erot cavity are split by the vibrating membrane. In such system the role of two resonances is played by upper and lower cavities. The membrane movement shift the cavity resonance frequencies in opposite directions with the same magnitude, $g_c/g_b=-1$. In the original setup the ratio $\Omega/g$ is very small, $\sim 10^{-4}$. However, as discussed  in Refs.~\cite{Wu_2013,Sankey2010},
it might be possible to use either a smaller membrane or  higher-order mechanical flexural modes of the membrane and  increase this ratio up to unity. 

Finally, a very different but promising platform is the ``picocavity optomechanics'', proposed in Refs.~\cite{Benz2016,Lombardi2018}. In such structures internal molecule vibrations interact with the localized plasmon modes. The vibration frequency is quite high, on the order of $30~$THz. Thus, we  can propose a setup where the molecule interacts with a plasmonic dimer as shown in Fig.~\ref{fig:structures}(d). The coupling between two modes of the dimer can be widely tuned by controlling the size and shape of the two nanoparticles and can be on the same order as the vibration frequency, $g\sim \Omega$.

\section{Theoretical approach}\label{sec:model}
We now proceed to the theoretical analysis of the light-matter coupling in the structure shown in Fig.~\ref{fig:1}. It is described by the following Hamiltonian:  
\begin{equation}
    \hat{H}=\hbar(\omega_c-g_c \hat{x})\hat{c}^\dagger\hat{c}+\hbar(\omega_b-g_b \hat{x})\hat{b}^\dagger\hat{b}+\hbar g(\hat{c}^\dagger \hat{b}+\hat{b}^\dagger \hat{c})\:.
\end{equation}
Here, $\omega_c$ is the cavity resonance frequency, $\omega_b$ is the exciton resonance frequency, $\hat{c}$ and $\hat{b}$ are the corresponding annihilation operators, and $g$ is the photon-exciton coupling strength. The operator $\hat{x}$ describes the mechanical displacement and $g_c$, $g_b$ are the corresponding optomechanical coupling constants. The operators $\hat{c}$ and $\hat{b}$ satisfy the Heisenberg equations of motion.
\begin{align}
    \dot{\hat{c}}&=\frac{1}{i\hbar}[\hat{c},\hat{H}]-\frac{\gamma_c}{2}\hat{c}-\sqrt{\gamma_c}\hat{a}_{\text{in}}\:,\label{eq:c}\\
     \dot{\hat{b}}&=\frac{1}{i\hbar}[\hat{b},\hat{H}]-\frac{\gamma_b}{2}\hat{b}\nonumber\:,
\end{align}
where $\gamma_c,\gamma_b$ are the cavity and exciton decay rates, respectively, and the operator $\hat{a}_{\rm in}$ describes the resonant optical pumping. From now on, we restrict ourselves to the semiclassical approximation, so that the quantum-mechanical operators can be replaced by their expectation values. In this case the system of equations Eq.~\eqref{eq:c} reduces to the following equations
\begin{align}\label{eq:c1}
    \dot{c}&=-i(\omega_c-g_c x)c-i g b -\frac{\gamma_c}{2} c-\sqrt{\frac{\gamma_c P_{\text{in}}}{\hbar\omega}}e^{-i\omega t}\:,\\ \nonumber
    \dot{b}&=-i(\omega_b-g_b x)b-i g c -\frac{\gamma_b}{2} b\:.
    \end{align}
Here, $P_{\rm in}$ is the pump power and $\omega$ is the pump frequency. The mechanical displacement is found from the Newton equation of motion
\begin{equation}\label{eq:x}
    \ddot{x}=\frac{\hbar g_c}{m}|c|^2+\frac{\hbar g_b}{m}|b|^2-\Omega^2 x-\Gamma  \dot{x}\:,
\end{equation}
where $\Omega$ is the mechanical frequency,  $\Gamma$ is the mechanical damping, $m$ is the mass of the oscillator. The terms $\propto |c|^2$
and  $\propto |b|^2$ describe the effect of the cavity and exciton on the oscillator motion, i.e. the radiation pressure force and the analogous exciton pressure force found as $F_{\rm rad}=-\partial H/\partial x$.

The system of differential equations Eq.~\eqref{eq:c1}--Eq.~\eqref{eq:x} can be solved numerically to find the motion of coupled light, exciton and phonons. However, it is also instructive to consider an approximate technique to describe the optomechanical damping and amplification. To this end we generalize the methodology from Refs.~\cite{Wu_2013}.
Within this approach the oscillator motion is described by the following ansatz:
\begin{equation}\label{eq:ansatz}
x=A\cos \Omega t\:,
\end{equation}
where we neglect the modification of the oscillator frequency and equilibrium position in the presence of radiation.
The optomechanical damping is determined by the average power of the radiation pressure. Namely, we multiply Eq.~\eqref{eq:x} by $\dot x$ and average  over the period of oscillations $2\pi/\Omega$ to find 
the optomechanical damping rate
\begin{equation}\label{eq:Gamma-opt}
    \Gamma_{\text{opt}}=-\frac{\hbar\langle (g_c|c|^2+g_b|b|^2)\dot{x}\rangle}{m\langle\dot{x}^2 \rangle}\:.
\end{equation}
In order to calculate the average we substitute the ansatz Eq.~\eqref{eq:ansatz} into Eq.~\eqref{eq:c1} and seek for the solution for $c(t)$, $b(t)$  in the form of Fourier series
\begin{equation}\label{eq:Fourier}
c=e^{-i\omega t}\sum_{n=-\infty}^
\infty c_n e^{i n\Omega t},\quad b=e^{-i\omega t}\sum_{n=-\infty}^
\infty b_n e^{i n\Omega t}\:.
\end{equation}
This yields the system of linear equations for the Fourier expansion coefficients $c_n$ and $b_n$. The system can be solved numerically by truncating the Fourier series.
Once the coefficients $c_n$ and $b_n$ are calculated, the optomechanical damping rate can be found from Eq.~\eqref{eq:Gamma-opt} as 
\begin{equation}\label{eq:Gammaopt}
       \Gamma_{\text{opt}}=-\frac{2\hbar}{\Omega Am}\Im \sum_n(g_c  c_n^*c_{n+1}+g_b b_n^* b_{n+1})\:.
\end{equation}

The Hamiltonian of a polaritonic system without optomecanical coupling can be written as 
\begin{equation}
    \hat{H}_0=\begin{pmatrix}
    \omega_c-\rmi \frac{\gamma_c}{2}&g\\g
    & \omega_b-\rmi \frac{\gamma_b}{2}
\end{pmatrix}\:.
\end{equation}
This Hamiltonian can be diagonalized using lower polariton $\ket{LP}$ and upper polariton $\ket{UP}$ eigenstates. Then the Green's function of this system $\hat{G}(\omega)=(\hat{H}_0-\omega \mathbb{1})^{-1}$ reads
\begin{equation}
    \hat{G}(\omega)=\frac{1}{\omega_{LP}-\omega}\ket{LP}\bra{LP}+\frac{1}{\omega_{UP}-\omega}\ket{UP}\bra{UP}\:,
\end{equation}
where $\omega_{\rm LP},\omega_{\rm UP}$ are the frequencies of lower and upper polariton, respectively. We also define the matrix responsible for the optomechanical coupling
\begin{equation}
    \hat{\mathcal G}=\begin{pmatrix}
    g_c & 0\\
    0 & g_b
    \end{pmatrix}\:. 
\end{equation}
Then optomechanical damping  $\Gamma_{\rm opt}$ Eq.~\eqref{eq:Gamma-opt} in the low oscillation amplitude limit reads 
\begin{align}\label{eq:Gamma-opt-an}
    \Gamma_{\text{opt}}&=\frac{\gamma_c P_{\text{in}}}{m\Omega\omega_c}\\&\times\Im\begin{pmatrix} 1 & 0\end{pmatrix} \hat{G}^\dagger(\omega)  \hat{\mathcal G}\left(\hat{G}^\dagger(\omega_{\rm S})+\hat{G}(\omega_{\rm aS})\right) \hat{\mathcal G} \hat{G}(\omega) \begin{pmatrix} 1 \\ 0 \end{pmatrix}\:.\nonumber
\end{align}
Here, $\omega_{\rm S}=\omega-\Omega$ and $\omega_{\rm aS}=\omega+\Omega$ are the Stokes and anti-Stokes frequencies.

\begin{figure}[t!]
\includegraphics[width=0.4\textwidth]{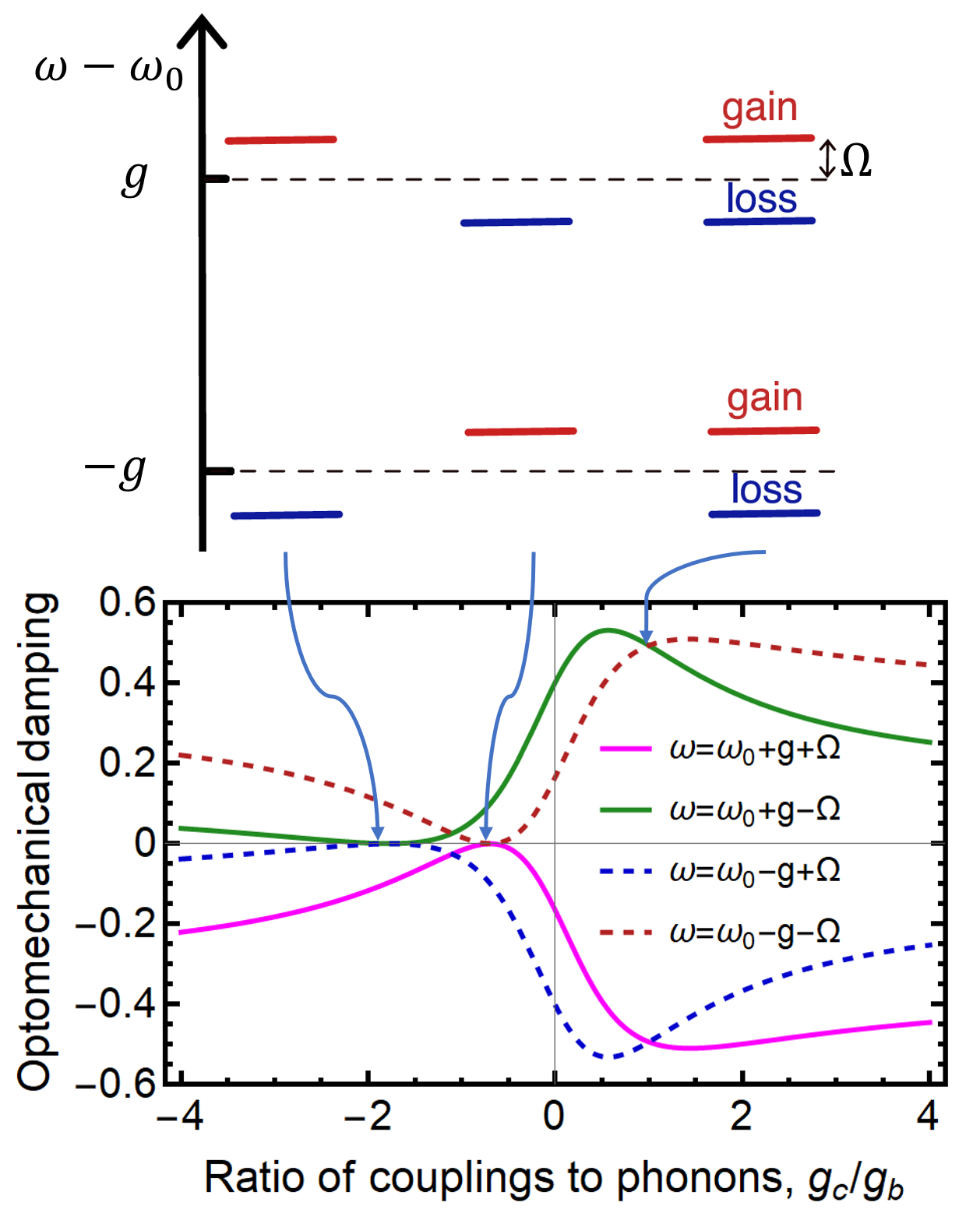}
\caption{Dependence of the optomechanical damping on the ratio of photon-phonon and exciton-phonon coupling constants $g_c/g_b$. Four curves correspond to different values of the pump frequency $\omega$, indicated on graph. Upper panel schematically illustrates spectra of optomechanical damping for three characteristic values of $g_c/g_b$ with red/blue lines illustrating amplification/damping for anti-Stokes and Stokes resonances of upper and lower polaritons. Calculation has been performed for the following set of parameters: $g=2.3\Omega$, $\gamma_c=\gamma_b=0.2 \Omega$. Optomechanical damping $\Gamma_{\rm opt}$ is measured in the units of $\mathcal{P}=P_{\rm in}(g_c^2+g_b^2)/(m\Omega^3\omega_c)$.
}\label{fig:2}
\end{figure}
\begin{figure}[t!]
\includegraphics[width=0.47\textwidth]{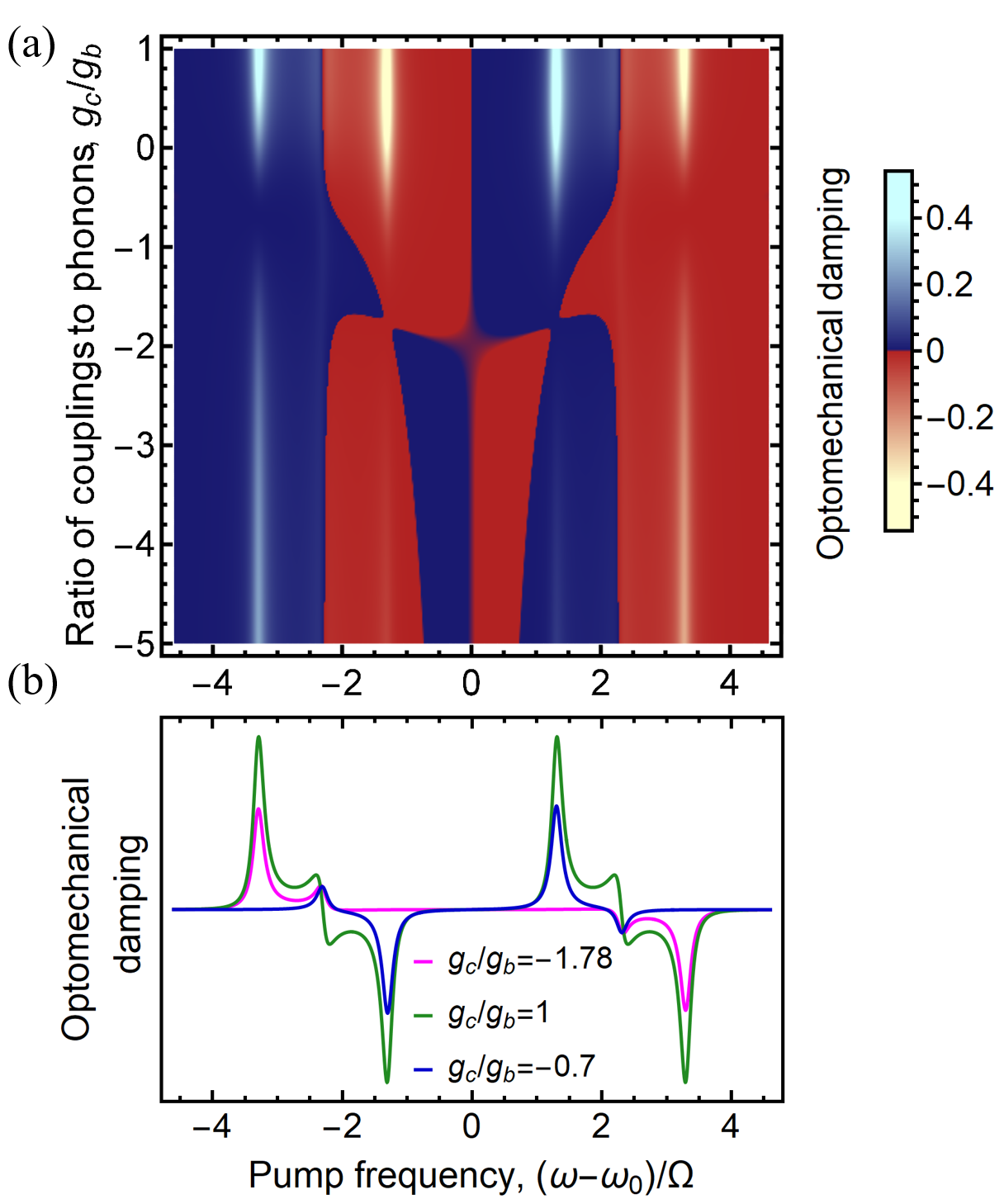}
\caption{
(a) Density plot of the optomechanical damping, as a function of the pump frequency $\omega$ and the ratio of the photon-phonon and exciton-phonon coupling constants coupling constants $g_c/g_b$. Panel (b) shows the damping spectra obtained as the cross sections of the density plot for three  values of $g_c/g_b$, indicated on graph. The calculation parameters are the same as in Fig.~\ref{fig:2}.
}\label{fig:3}
\end{figure}
\begin{figure}[tb]
\includegraphics[width=0.48\textwidth]{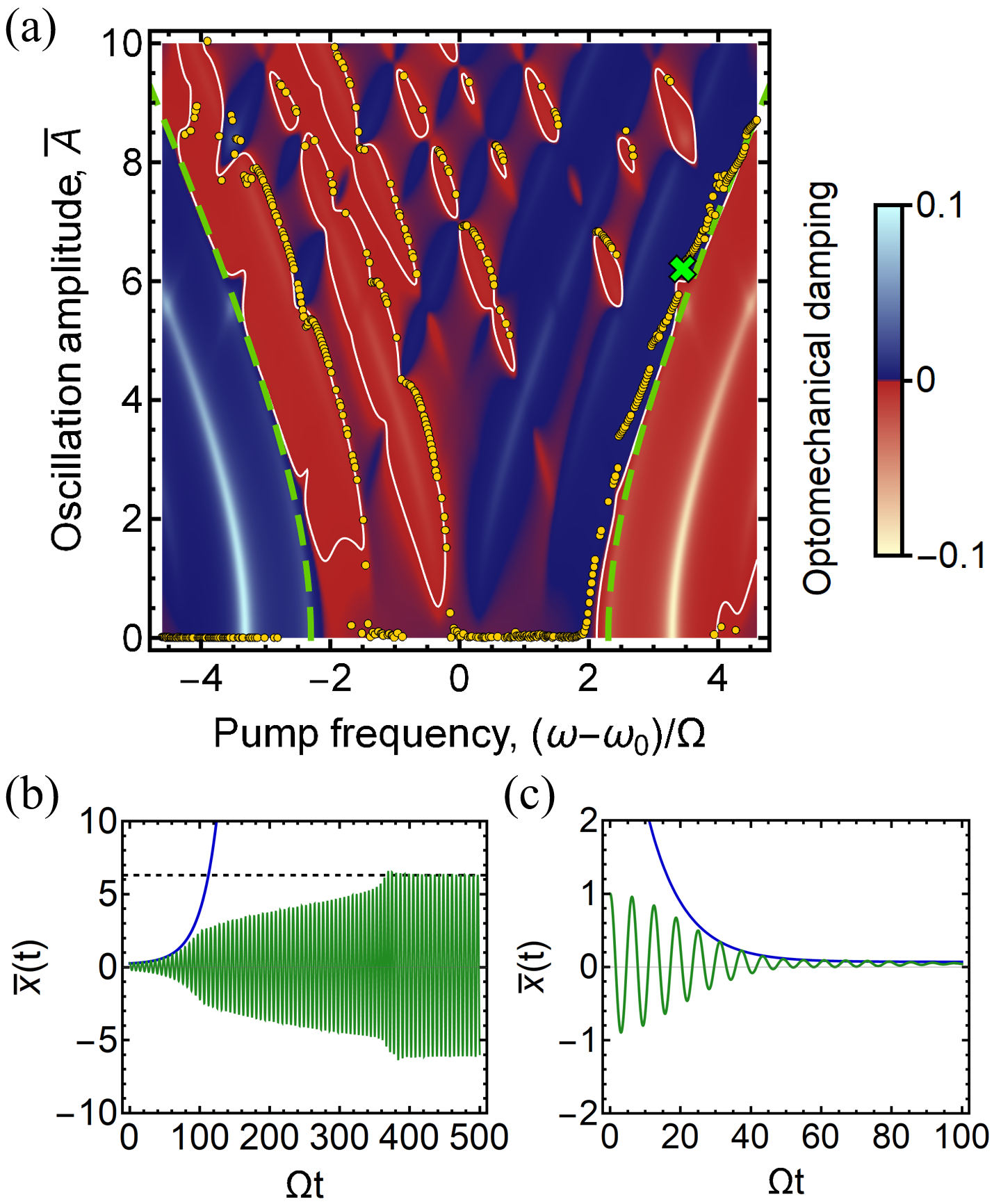}
\caption{(a) Density plot of the  optomechanical damping, as a function of the pump frequency $\omega$ and the oscillation amplitude $\overline{A}=A\sqrt{g_c^2+g_b^2}/\Omega$. White contour lines display attractors $\Gamma_{\rm opt}=-\Gamma$ for $\Gamma=10^{-3} \mathcal{P}$ ($\mathcal{P}/\Omega=1$). The ratio of photon-phonon and exciton-phonon coupling constants is $g_c/g_b=-1.78$. Yellow dots show the results of numerical simulation of
the dynamical equations of motion \eqref{eq:c1},\eqref{eq:x}. Panels (b) and (c) show the time evolution of the mechanical displacement for pump frequencies $\omega-\omega_0=3.4\Omega$ and $\omega-\omega_0=-3.3\Omega$, respectively. Green dashed lines show the approximate analytical equation Eq.~\eqref{eq:anticross} for the polariton resonances.
 Other calculation parameters are the same as in Fig.~\ref{fig:2}.
}\label{fig:4}
\end{figure}
\section{Asymmetric optomechanical damping spectra}\label{sec:results}
We now proceed to the discussion of the dependence of optomechanical damping and amplification on the ratio of the coupling constants $g_c/g_b$. We consider the strong coupling regime with $g\gg \gamma_c, \gamma_b$, and cavity and exciton resonances tuned to each other, $\omega_c=\omega_b\equiv \omega_0$. In this case the upper and lower polariton resonance are well defined and have the eigenfrequencies $\omega_0\pm g$.
The results of calculation are shown in Figs.~\ref{fig:2},\ref{fig:3}. Figure~\ref{fig:2} presents optomechanical damping for the pumping frequency $\omega$ tuned to the Stokes or the anti-Stokes frequency of upper (solid curves) and lower (dashed curves) polariton resonance,    $\omega_0\pm g \pm \Omega$.  Upper panel schematically illustrates  optomechanical damping for three characteristic values of $g_c/g_b$ at these four frequencies. The calculation results strongly depend on the ratio of the coupling constants. Namely, for $g_c=g_b$ the spectra manifest mirror-symmetric damping and amplification around each of the two polariton resonances. On the other hand, for negative values of $g_c/g_b$ the spectra become strongly asymmetric. Specifically, for $g_c/g_b\approx -1$, the amplification is suppressed for upper polariton resonance, and the damping is suppressed for lower polariton resonances. The situation when $g_c/g_b\approx -2$ corresponds to the opposite case, when only Stokes upper polariton resonances and anti-Stokes lower polariton resonances survive in the spectrum.

The same spectral asymmetry can be seen from the density plot of the optomechanical damping, as a function of the pump frequency $\omega$ and the ratio of the coupling constants $g_c/g_b$ shown in Fig.~\ref{fig:3}. While for $g_c/g_b>0$ there exist two distinct regions of damping (blue color) alternating with two regions of amplification (red color), the spectral shape is strongly modified for $g_c/g_b<-1$. Namely, the spectra become asymmetric and acquire three regions of damping and three regions of amplification. The same can be seen from  the damping spectra shown in the lower panel of Fig.~\ref{fig:3}.

The observed  asymmetry of the spectra of optomechanical damping/amplification depending on $g_c/g_b$ is the main result of our work. We now proceed to the analytical description of this effect.

It can be seen from  Eq.~\eqref{eq:Gamma-opt-an} that at low oscillation amplitudes the damping has  several terms due to the interference of lower and upper polaritons. The asymmetry described above manifests itself due to the strong dependence of the interference terms on the parameters of the system. In case $\omega_c=\omega_b$ and $\gamma_c=\gamma_b$, eigenstates of polaritonic system are $\ket{LP}=\frac{1}{\sqrt{2}}(-1,1)^T$ and $\ket{UP}=\frac{1}{\sqrt{2}}(1,1)^T$. This special case allows us to obtain a relatively simple expression for the optomechanical damping. We are especially interested in the optomechanical damping for pump frequencies at the Stokes and anti-Stokes resonances of the polaritons, $\omega-\omega_0=\pm g \pm \Omega$, shown also in Fig.~\ref{fig:2}. Keeping only the resonant terms in the spectrum, we find the proportionality coefficient. For example, for the pump frequency $\omega-\omega_0=g+\Omega$ we find
\begin{equation}
        \Gamma_{\text{opt}}\propto f(g+\Omega)= \left(g_b+\left(1+\frac{\Omega}{g}\right)g_c\right)^2\:.
\end{equation}
From here we see that the parameter responsible for the spectrum asymmetry is $\Omega/g$. For $\Omega/g\ll1$ we have $f(g+\Omega)\approx f(g-\Omega)$. In this case upper and lower polariton resonances can be considered separately and spectrum has the same mirror-symmetric damping and amplification on the lower and upper polaritons for all values $g_c/g_b$. However, for  $\Omega/g\sim 1$ the damping strongly depends on the ratio of the coupling constants. We find the extremum of the expression $f(g+\Omega)/f(g-\Omega)$ to determine the values $g_c/g_b$ at which the maximum asymmetry is observed 
\begin{equation}
    \frac{g_c}{g_b}=-\frac{1}{1\pm \Omega/g}\:.
\end{equation}
The resulting two values of $g_c/g_b$ correspond to the blue and magenta  curves in Fig.~\ref{fig:3}~(b).

As shown in Fig.~\ref{fig:2} we can achieve an increase in damping or amplification on one of the polaritons by choosing ratio of $g_c/g_b$. These ratios can be found from the maximum of the expression $f(g\pm\Omega)/(g_c^2+g_b^2)$:
\begin{equation}
    \frac{g_c}{g_b}=1\pm\frac{\Omega}{g}\:.
\end{equation}
\section{Nonlinear regime}\label{sec:nonlinear}

The previously considered regime  corresponds to  small amplitude oscillations. Now we proceed to describe the behavior of the system in a strongly nonlinear regime when the oscillation amplitude is large. The optomechanical gain  $\Gamma_{\rm opt}$ calculated following Eq.~\eqref{eq:Gammaopt} is presented in Fig.~\ref{fig:4}~(a). Possible attractors of the oscillation amplitude $\overline{A}$ shown by white contour lines correspond to zero total damping ($\Gamma_{\rm opt}=-\Gamma$), when the optomechanical gain is balanced by friction. Stable attractors are those where $\Gamma_{\rm opt}$ increases for increasing $\overline{A}$~\cite{Marquardt2006}. The yellow dots show the long-time limit of the amplitude in the numerically exact solution of the equations of motion, Eqs.~\eqref{eq:c1},\eqref{eq:x}. It can be seen that these dots are in good agreement with the stable attractors. 

The calculation shows that the upper and lower polariton resonances move apart with increasing amplitude. This behavior can be explained by averaging  the eigenfrequencies of  the optomechanical system, obtained for a fixed value of the coordinate $x$, 
\begin{equation}
    \omega_{\pm}(t)=\frac{g_c+g_b}{2}x(t)\pm \sqrt{g^2+\left(\frac{g_c-g_b}{2}\right)^2x^2(t)}\:,
\end{equation}
over the over the oscillation period $2\pi/\Omega$. 
Effective resonance frequencies are described by the following equation
\begin{gather}\label{eq:anticross}
    \langle\omega_{\pm}(t)\rangle=\pm \frac{2\sqrt{g^2+\alpha^2}}{\pi}E\left(\frac{\pi}{2},\frac{\alpha}{\sqrt{g^2+\alpha^2}}\right)\:,\\ \nonumber
    \alpha=\frac{\Omega\overline{A}}{2}\frac{g_c-g_b}{\sqrt{g_c^2+g_b^2}}\:,
\end{gather}
where $E(\pi/2,\alpha/\sqrt{g^2+\alpha^2})$ is the complete elliptic integral of the second kind, shown by green dashed lines in Fig.~\ref{fig:4}(a). Another effect, seen in Fig.~\ref{fig:4}(a) is the presence of many sidebands that are shifted from the main ones by an integer number of phonon frequencies $\Omega$. These sidebands correspond to multi-phonon absorption or emission~\cite{Wu_2013}.

The calculation demonstrates, that the spectral asymmetry of the optomechanical gain, discussed previosly in the regime of weak oscillations  [Figs.~\ref{fig:2},\ref{fig:3}], persists in the nonlinear regime.  For the parameters of Fig.~\ref{fig:4}, the optomechanical  damping is enhanced near the lower polariton resonance, while  there exists opomechanical amplification near the upper polariton resonance. Figures~\ref{fig:4}~(b) and (c) show the time evolution of the mechanical displacement for pump frequencies corresponding to strong amplification and damping, respectively. In the first case, the amplitude increases exponentially at the beginning and after $\Omega t\approx400$ approaches a stationary value ($\overline{A}\approx6.2$), which is marked on the density plot with the green cross. In the second case, the oscillations simply decay exponentially.

\section{Summary}\label{sec:summary}
We have calculated optomechanical damping and amplification for a model, where a single vibration mode is simultaneously coupled with two resonant oscillators. Such semiclassical model describes      confined phonons, excitons and light interacting in a microcavity and forming excitonic polaritons. We demonstrate, that in case when the vibration frequency is not much smaller than the Rabi splitting between the upper and lower polaritonic modes,  the optomechanical damping and amplification spectra become strongly asymmetric and very sensitive to the nature of polariton-photon coupling. Namely, the spectral shape significantly depends on  the ratio of the exciton-phonon and the photon-phonon coupling constants.  We have obtained an analytical expression for the damping/amplification  spectra for low vibration amplitudes that reveals the key role of interference between these two couplings in the spectral asymmetry. By tuning the ratio of the couplings it is possible to realize selective  optomechanical gain, existing either only for lower polaritonic resonance or only for upper polaritonic resonance.  Our calculation demonstrates, that the spectral asymmetry, revealed for low vibrations amplitude, also exists for larger amplitudes, where  optomechanical nonlinearity plays an important role. The nonlinear attractor diagram of the resulting phonon lasing oscillations is  strongly  asymmetric  and very sensitive to the origin of the  polariton-photon coupling. Our results apply to different optomechanical systems,  where several resonant oscillators, such as photons, plasmons, or excitons interact with the same vibration mode.

\acknowledgements
We acknowledge useful discussions with A.V.~Poshakinskiy. This work has been funded by the Russian Science Foundation Project No.~20-42-04405. 
%

\end{document}